\documentclass[
  ,draft            % use draft while you are working on the paper
  ]{aipproc}

\layoutstyle{6x9}

\begin{document}

\title{A minor-merger origin for inner disks and rings in early-type galaxies}


\author{M.~C.~Eliche-Moral}{
  address={Dpto.~de Astrof\'{\i}sica, Universidad Complutense de Madrid, 28040, Madrid, Spain}
}

\author{A.~C.~Gonz\'{a}lez-Garc\'{\i}a}{
  address={Dept.~F\'{\i}sica Te\'{o}rica, Universidad Aut\'{o}noma de Madrid, 28049, Madrid, Spain}
}

\author{M.~Balcells}{
  address={Instituto de Astrof\'{\i}sica de Canarias, 38200 La Laguna, Tenerife, Spain},
  altaddress={Isaac Newton Group of Telescopes, Aptdo.~321, 38700 Santa Cruz de La Palma, Spain}
}

\author{J.~A.~L.~Aguerri}{
  address={Instituto de Astrof\'{\i}sica de Canarias, 38200 La Laguna, Tenerife, Spain}
}

\author{J.~Gallego}{
  address={Dpto.~de Astrof\'{\i}sica, Universidad Complutense de Madrid, 28040, Madrid, Spain}
}

\author{J.~Zamorano}{
  address={Dpto.~de Astrof\'{\i}sica, Universidad Complutense de Madrid, 28040, Madrid, Spain}
}

\begin{abstract}
Nuclear disks and rings are frequent galaxy substructures, for a wide range of morphological types \citep[from S0 to Sc, see e.g.,][]{2010AIPC....M,2010AIPC....S}. We have investigated the possible minor-merger origin of inner disks and rings in spiral galaxies through collisionless N-body simulations. The models confirm that minor mergers can drive the formation of thin, kinematically-cold structures in the center of galaxies out of satellite material, without requiring the previous formation of a bar. Satellite core particles tend to be deposited in circular orbits in the central potential, due to the strong circularization experienced by the satellite orbit through dynamical friction. The material of the satellite core reaches the remnant center if satellites are dense or massive, building up a thin inner disk; whereas it is fully disrupted before reaching the center in the case of low-mass satellites, creating an inner ring instead. 
\end{abstract}

\maketitle
%%%%%%%%%%%%%%%%%%%%%%%%%%%%%%%%%%%%%%%%%%%%
%% MAINMATTER
%%%%%%%%%%%%%%%%%%%%%%%%%%%%%%%%%%%%%%%%%%%%

\section{Models and Results}
\begin{figure}
  \includegraphics[width=0.48\textwidth]{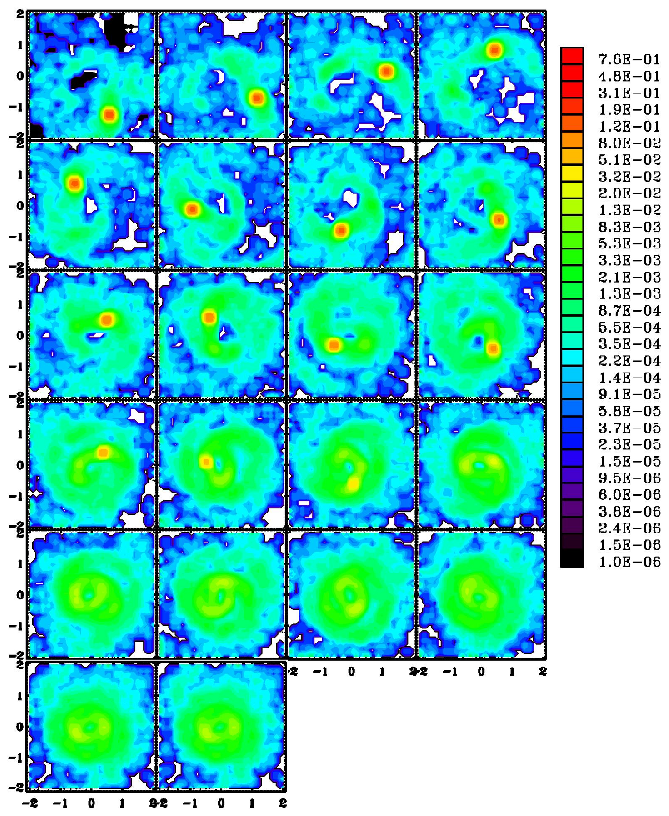}
  \includegraphics[width=0.48\textwidth]{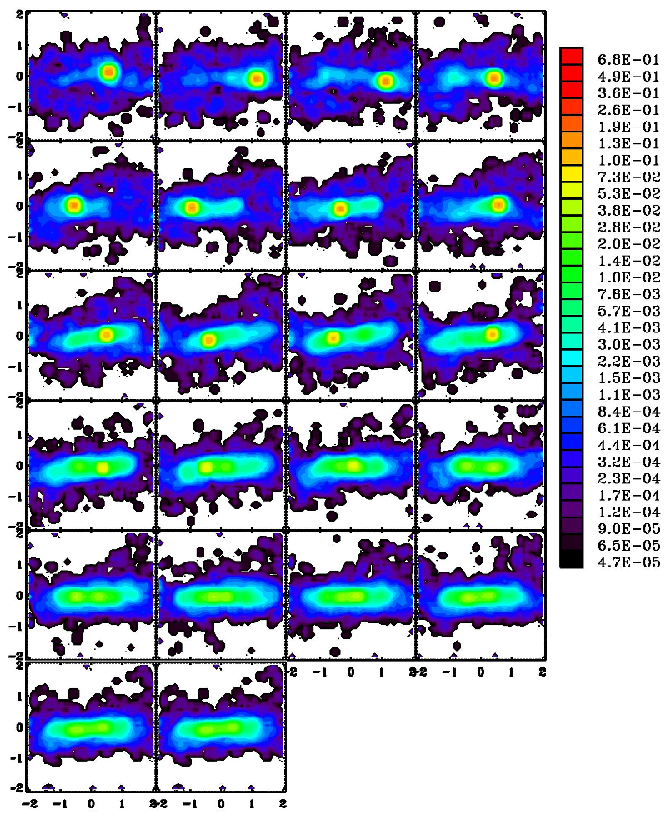}
  \caption{Evolution of the luminous surface density of the satellite bulge material, during the last moments of one of the experiments. Face-on and edge-on views centered in the primary galaxy are plotted. Bulge satellite material is deposited significantly since the second pericenter passage in the remnant center, building up an inner ring.}\label{fig:fig1}
\end{figure}

\vspace{-0.3cm}
Inner rings and disks have been traditionally associated to bars, except for those cases of extremely inclined geometries with respect to the host disk, thought to have been formed through face-on mergers \citep[][]{2009NewAR..53..169M}. However, recent observations pose that these inner structures, primarily found in Sa-Sbc galaxies, are not predominantly associated to bars \citep{2006A&A...448..489K,2009arXiv0908.0272C}. Disk dynamical resonances and oval distortions induced by minor mergers have also been proposed as formation mechanisms \citep{2006AJ....131.1336S,2009arXiv0910.0589B}. We have studied if minor mergers can really drive the formation of thin, dynamically-cold components in the center of disk galaxies, using collisionless N-body simulations. We have completed the set of minor merger experiments described in \citet{2006A&A...457...91E}, using different orbital configurations (direct and retrograde, small and large pericenter distances), mass ratios (1:18, 1:9, 1:6), and initial bulge-to-disk ratios of the primary galaxy ($B/D=0.1$, 0.5). While the satellite disk deposits its mass onto the remaining primary disk during the merger, the particles from the satellite bulge settle into a vertically-thin structure in the center of the remnant, forming nuclear rings or disks that span from 0.1-0.5 disk scale lengths (see Fig.\,\ref{fig:fig1}). No significant bar-distortions are induced in the disks by the interactions. Specific momentum-energy plots reveal that these structures are kinematically-cold, due to the strong orbital circularization experienced by the satellite core through dynamical friction during the decay. Although the surface density of these structures is lower than that of the underlying disk (making their detection difficult), they could become detectable having into account the star formation that minor mergers usually entail \citep{2009MNRAS.394.1713K}. Our models suggest that minor mergers could be relevant drivers for the formation of inner cold structures in non-barred, early-type spirals.

\vspace{0.2cm}
{\footnotesize Supported by the Spanish Programa Nacional de Astronom\'{\i}a y Astrof\'{\i}sica (project AYA2006-02358), by the Madrid Regional Government through the ASTRID Project (S0505/ESP-0361) for development and exploitation of astronomical instrumentation (http://www.astrid-cm.org/), and by the Spanish MICINN under the Consolider-Ingenio 2010 Program grant CSD2006-00070: "First Science with the GTC" (http://www.iac.es/consolider-ingenio-gtc/).\/}
\vspace{-0.5cm}
%%%%%%%%%%%%%%%%%%%%%%%%%%%%%%%%%%%%%%%%%%%%
%% Sample figure:
%%
%% The option [height=...] scales the picture to the given height,
%% without it it would be printed at its nominal size
%%%%%%%%%%%%%%%%%%%%%%%%%%%%%%%%%%%%%%%%%%%%

%%%%%%%%%%%%%%%%%%%%%%%%%%%%%%%%%%%%%%%%%%%%%%%%
%% BACKMATTER
%%%%%%%%%%%%%%%%%%%%%%%%%%%%%%%%%%%%%%%%%%%%%%%%

%\begin{theacknowledgments}
%{\footnotesize
%Supported by the Spanish Programa Nacional de Astronom\'{\i}a y Astrof\'{\i}sica (project AYA2006-02358), by the Madrid Regional Government through the ASTRID Project (S0505/ESP-0361) for development and exploitation of astronomical instrumentation (http://www.astrid-cm.org/), and by the Spanish MICINN under the Consolider-Ingenio 2010 Program grant CSD2006-00070: "First Science with the GTC" (http://www.iac.es/consolider-ingenio-gtc/).
%\/}
%\end{theacknowledgments}



\bibliographystyle{aa}   % if natbib is available

\bibliography{mybib.bib}


\end{document}